# Vector Electrometry in a Wide-Gap Semiconductor Device Using a Spin Ensemble Quantum Sensor

Yang, Bang[1], Takuya Murooka[1], Kwangsoo Kim[1], Hiromitsu Kato[2], Toshiharu Makino,[2] Masahiko Ogura[2], Satoshi Yamasaki[2], Amir Yacoby[3,4,5], Mutsuko Hatano[1], and Takayuki Iwasaki[1,*]

[1]Department of Electrical and Electronic Engineering, School of Engineering, Tokyo Institute of Technology, Meguro, Tokyo 152-8552, Japan

[2]Advanced Power Electronics Research Center, National Institute of Advanced Industrial Science and Technology, Tsukuba, Ibaraki 305-8568, Japan

[3]Department of Physics, Harvard University, 17 Oxford Street, Cambridge, Massachusetts 02138, USA

[4]John A. Paulson School of Engineering and Applied Sciences, Harvard University, Cambridge, Massachusetts 02138, USA

[5]Tokyo Tech World Research Hub Initiative (WRHI), School of Engineering, Tokyo Institute of Technology, Meguro, Tokyo 152-8552, Japan

e-mail: iwasaki.t.aj@m.titech.ac.jp

**Abstract**

Nitrogen-vacancy (NV) centers in diamond work as a quantum electrometer. Using an ensemble state of NV centers, we propose vector electrometry and demonstrate measurements in a diamond electronic device. A transverse electric field applied to the N-V axis under a high voltage was measured while applying a transverse magnetic field. The response of the energy level shift against the electric field was significantly enhanced compared with that against an axial magnetic field. Repeating the measurement of the transverse electric field for multiple N-V axes, we obtained the components of the electric field generated in the device.

The electric field is a driving force of systems in a variety of fields such as electronic devices, electrochemical reactions, and cell membranes. Detecting the electric field inside such systems is a challenging task. High-resolution electrometry can be performed using a scanning probe microscope [1,2] with the capability of making vector measurements [3]. However, these methods only give information at the vicinity of the surface. Recently, electron spins in solid-state materials such as nitrogen-vacancy (NV) centers in diamond [4–18] and Si vacancies in SiC [19,20] have emerged as electrometers. The NV center is a complex defect composed of a pair of nitrogen atom and a neighboring vacancy in diamond [21]; it also works as a quantum sensor for magnetic fields [22–24], temperature [25–29], and pressure [30]. NV-based electrometry was first reported by Van Oort et al [4], in which they determined the electric dipole moments of the NV center. Dolde et al [5] reported the measurement of axial and transverse electric fields using a single NV center. However, the axial electric susceptibility factor is much lower than the transverse parameter. Thus, usually, the electric field transverse to one NV alignment is measured for electrometry [14,15], hindering the gathering of vector information on the electric field, which would provide deep insight into the operation of the systems.

Vector electrometry was proposed using an ensemble state of NV centers with multiple N-V axes [14]. In the proposed method, a large axial magnetic field is applied to multiple N-V axes, which significantly lowers the shift of the energy level against the electric field. In this study, we propose a method to overcome this issue. Multiple N-V axes are selected as target NV centers. A transverse magnetic field is applied to one of the target NV alignments, and then, an electric field is also generated in the system, giving rise to an electric field transverse to the target NV. By changing the target N-V axis and by repeating the process, the components of the electric field can be estimated using the information from the multiple N-V axes. Here, we demonstrate dc vector electrometry in a diamond electronic device using ensemble NV centers. Diamond power devices are promising candidates for next-generation low-loss power electronics because of their wide band gap, high carrier mobilities, and high thermal conductance [31]. The wide band gap of diamond makes it possible to sustain a high electric field. However, most of the reported diamond devices do not attain an ideal electric field [32], requiring direct detection of the electric field. Furthermore, diamond devices work as a platform for electrical control and sensing for quantum technologies [33–40], in which the internal electric field plays an important role in the operation of the devices. NV electrometry can provide direct information about the electric field inside the devices. Different from our previous work using single NV centers [13], here, the spatial distribution of the electric field was observed with an ensemble of NV centers.

**I. Principle of vector electrometry**

The important characteristic of the NV center for vector electrometry is the four possible crystallographic alignments of the N-V axis in the crystal (double alignments exist when considering the N-V and V-N directions). Since an N-V axis takes one alignment of [111], [$\bar{1}1\bar{1}$], [1$\bar{1}\bar{1}$], or [$\bar{1}\bar{1}$1], as shown in Fig. 1a, ensemble NV centers include all alignments. The V-N direction of NV A is consistent with the z-axis, and that of NV B is on the xz plane in the lab frame shown in Fig. 1d. Note that for analysis of the electric field, we use a different coordinate for each NV alignment from that in the lab frame. For each NV coordinate, the z-axis is defined as the direction from the N atom to vacancy, the x-axis is perpendicular to the z-axis, and the xz plane includes one of the carbon atoms neighboring the vacancy [11]. In this study, the (x, z) coordinate for NV A is the ($\bar{1}\bar{1}1$, $\bar{1}\bar{1}2$) direction. Similarly, the (x, z) coordinate for NV B corresponds to the (112, 11$\bar{1}$) direction.

We can perform vector electrometry by using multiple NV axes, similar to vector magnetometry [41,42]. To recognize the relationship between the NV axes and optically detected magnetic resonance (ODMR) signals, an arbitrary directed magnetic field is applied for vector magnetometry. However, for electrometry, the axial magnetic field significantly lowers the shift of the ODMR dips resulting from an electric field (Fig. 3, Appendix A). This issue can be overcome using an approach of applying a transverse magnetic field to a NV axis (Fig. 1b) [14,15]. For instance, upon application of a transverse magnetic field to NV A, the corresponding ODMR signals remain close to the zero-field splitting ($D_{gs}$), while other NV axes (NV B-D) are largely split by axial magnetic fields. Thus, we can recognize the signals from NV A. Figure 1c shows an ODMR spectrum while applying a transverse magnetic field, along the x-axis in the lab frame, i.e., [$\bar{1}\bar{1}2$] direction, to NV A. The direction of the magnetic field was controlled by using three-axis electromagnets (Appendix B). The innermost two dips at ~2.87 GHz in the spectrum correspond to NV A, while other signals come from NV B-D.

Then, the NV A resonance splits further with the application of a transverse electric field (Fig. 1b). The splitting width of ODMR, W, of NV A is described in equation (1) [5,11] (see also Appendix A), taking $B_y$, $B_z$~0 in the NV A coordinates.

$$\frac{1}{2}W = \left[k_\perp^2 \Pi_\perp^2 - \frac{\gamma^2 B_\perp^2}{2D_{gs}} k_\perp \Pi_\perp \cos\phi_\Pi + \frac{\gamma^4 B_\perp^4}{4D_{gs}^2}\right]^{\frac{1}{2}} \quad (1)$$

where $\gamma$ denotes the gyromagnetic ratio, provided as 28 GHz/T, $B_\perp$ represents the transverse magnetic field to the NV axis, defined as $B_\perp = \sqrt{B_x^2 + B_y^2}$, and $k_\perp$ denotes the transverse electric susceptibility parameter, provided as 17 Hz/V/cm [4]. $\Pi_\perp$ is the effective electric field perpendicular to the NV axis, defined as $\mathbf{\Pi}_\perp = \mathbf{E}_\perp + \mathbf{\sigma}_\perp$, where $\mathbf{E}_\perp$ and $\mathbf{\sigma}_\perp$ are the transverse electric and strain fields, respectively. The magnitude of $\mathbf{\Pi}_\perp$ is given as $\Pi_\perp = \sqrt{\Pi_x^2 + \Pi_y^2}$, and $\tan\phi_\Pi = \Pi_y/\Pi_x$. It is found that the two NV orientations, i.e., N-V and V-N, take different splitting widths [11,14,43]. We expect that ion implantation and annealing reasonably lead to equal fabrication of the two NV orientations in the ensemble state. The difference in the resonance frequency is below 2.5 MHz under the measurement conditions in this study. Furthermore, inhomogeneous broadening of the ODMR signals will occur due to the variation of the electric field strength at the measurement region, hindering the observation of the small shift owing to the NV orientations. Thus, here, we analyze the experimentally obtained ODMR splitting as the average of the two orientations. According to equation (1), the splitting increases almost linearly with the electric field, as shown by the solid lines in Fig. 3. At an electric field of ~1.1 MV/cm, the split increase becomes larger than 30 MHz, compared with ~3 MHz for the case with an axial magnetic field (Appendix A). Therefore, the response to the electric field can be increased by a factor of 10 in this field range.

The abovementioned process is related to one NV axis. Vector electrometry can be achieved by repeating this process for multiple NV axes, i.e., applying a transverse magnetic field to a target NV axis and then applying the electric field in the system. Finally, the electric field generated in the device can be estimated from a system of equations.

$$E_\perp^{NV\,i} = |\mathbf{E} - (\mathbf{E} \cdot \mathbf{u}_{NV\,i})\mathbf{u}_{NV\,i}| \quad (i = A, B, C, D) \qquad (2)$$

where $E_\perp^{NV\,i}$ is the transverse electric field to each NV alignment, $\mathbf{E} = (E_x, E_y, E_z)$ in the lab frame, and $\mathbf{u}_{NV\,i}$ is the unit vector in each NV direction (Appendix C). In this study, we used two NV axes of NV A and NV B to obtain $E_x$ and $E_z$.

**II. Device structure and measurement setup**

We performed vector electrometry in a diamond device using ensemble NV centers. Figure 1d shows a vertical diamond p-i-n diode on a boron-doped ([B]= $1\times10^{17}$ cm$^{-3}$) (111) diamond single-crystal substrate. In terms of electrical properties, the device shows a high rectification ratio of ~$10^6$ at $\pm 10$ V (Appendix D). The detailed fabrication process of the device is described elsewhere [13,44]. The device has patterned heavily phosphorus-doped ([P]= $1\times10^{19}$ cm$^{-3}$) 350 nm-thick n$^+$-regions on top of a 5 μm-thick intrinsic layer. We take one edge of the n$^+$-region as the origin "O" for the xyz axes in the lab frame, as shown in Fig. 1d. The x and z axes in the lab frame are along the $[\bar{1}\bar{1}2]$ and $[\bar{1}\bar{1}\bar{1}]$ directions, respectively. The spacing between the n$^+$-regions is 10 μm. Nitrogen ion implantation was performed over the device with a dose of $1\times10^{12}$ cm$^{-2}$ and an acceleration energy of 350 keV at an elevated temperature of 600°C, giving rise to a projected depth of ~350 nm. Then, the device was annealed at 750°C for 30 min to promote the diffusion of the vacancies and form the NV centers. Thus, ensemble NV centers were formed in the i-layer 350 nm from the surface and at the interface between the n$^+$-regions and i-layer. The diamond substrate with the devices was placed in a vacuum chamber with a quartz window for optical access (Fig. 1e). The vacuum environment (~6×10$^{-3}$ Pa) was used to avoid discharge in air at a high voltage of 400 V. The NV centers were excited using a 532 nm green laser through an objective lens (numeric aperture: 0.95) placed on three-axis piezo scanners in the chamber, and their fluorescence was measured with a charge-coupled device (CCD) (Fig. 1f). Microwave (MW) radiation was applied though a Cu wire for electron spin resonance. All experiments in this study were performed at room temperature.

The arrows with dashed lines in the i-layer shown in Fig. 1d denote the expected electric field generated in the device upon the application of a high reverse voltage. Below the n$^+$-regions, the electric field in the z direction is dominant in the p-i-n diode. On the other hand, the electric field concentrates at the edge of the patterned n$^+$-regions and thus has components in both the x and z directions. At a high voltage of 400 V, the directions of $E_x$ and $E_z$ are positive at the measurement position (Appendix E). The length of the n$^+$-regions along the y-axis is 50 μm, and the electric field measurements were performed close to the y center. Thus, we assume no electric field in the y direction along the electrode, and the binning of the pixels in the y direction was performed for the electric field analysis.

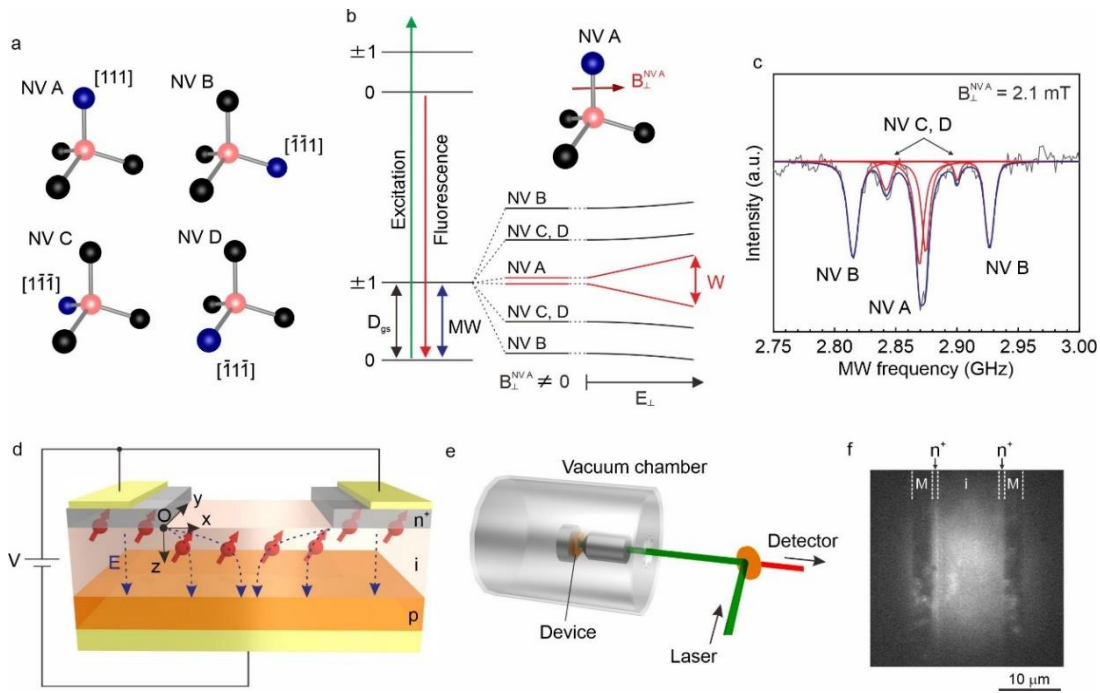

Figure 1. Vector electrometry using ensemble NV centers. (a) Four possible NV alignments. The blue, red, and black spheres represent N, V, and C atoms, respectively. (b) Energy level of the NV center. The energy splitting in the ground state is shown for the application of transverse magnetic and electric

fields to NV A. (c) ODMR spectrum under a transverse magnetic field of 2.1 mT to NV A. The gray curve represents the experimental data. The red and blue curves are the fits and envelope of the fitting curves, respectively. (d) Vertical diamond p-i-n diode on a (111) diamond substrate incorporating the NV centers. "O" denotes the origin of the xyz axes in the lab frame. The x and z axes are along $[\bar{1}\bar{1}2]$ and $[\bar{1}\bar{1}\bar{1}]$ directions, respectively. The dashed lines indicate expected electric fields generated in the i-layer of the device. Note that the NV centers shown as red objects were fabricated uniformly along the y direction, but they are omitted in this figure for the sake of clarity. (e) Measurement setup. The laser goes into the chamber along the z-axis in the lab frame as shown in panel (d). (f) CCD image of the fluorescence intensity in a device. M denotes cathode metal electrodes on top of the $n^+$-regions.

### III. Vector electrometry in a diamond electronic device

Before the application of the electric field, the magnitude of the transverse magnetic field applied to a target NV alignment (here, NV A) is estimated. The six resonant signals are observed in the ODMR spectrum shown in Fig. 1c. Since the magnetic field was applied along the x-axis in the lab frame, i.e., the $[\bar{1}\bar{1}2]$ direction, the outermost signals at 2.82 and 2.93 GHz correspond to NV B. Due to the symmetry of NV C and NV D, their energy levels are degenerate, appearing at the same frequencies of 2.84 and 2.90 GHz. The magnitude of the transverse magnetic field applied to NV A was estimated to be 2.1 mT from the splitting width of the NV B signals.

In the next step, we applied a reverse voltage of 400 V to the device while applying the transverse magnetic field to target NV A (Fig. 2a). Figure 2c shows the ODMR spectra depending on the x position in the device. The position at x=5 μm corresponds to the midpoint of the i-layer between the $n^+$-diamond regions, where the lowest electric field is expected along the x-axis. Approaching the edge of the $n^+$-region (x=0 μm or O point in Fig. 1d) should increase the electric field due to the field concentration. In the case of taking NV A as a target (Fig. 2c), a pair with a large splitting appears in the ODMR spectrum at x=2 μm, and its splitting becomes even larger at x=0 μm by sensing $E_x$, mentioned as $f_1^-$ and $f_1^+$. Another pair with a smaller splitting ($f_2^-$ and $f_2^+$) would come from NV A either at the $n^+$-i interface or at a small electric field position in the i-layer.

Then, in the same way, we applied a transverse magnetic field of 4.1 mT to the next target axis, NV B (Fig. 2b). The $E_z$ value is expected to be higher than $E_x$ in the p-i-n diode. Hence, a higher magnetic field was applied for the case of NV B to avoid overlapping the NV B signals with those from other NV axes upon application of the electric field. Upon the application of 400 V, the ODMR spectra at x=5 and 2 μm show similar signals (Fig. 2d). At x=0 μm, however, the splitting width rapidly increases, and multiple pairs are observed. Compared with $E_x$, the large $E_z$ component is generated in the vertical device structure. Thus, the outermost signals ($f_1^-$ and $f_1^+$) are thought to originate from NV B at the $n^+$-i interface. The pair with the second largest splitting ($f_2^-$ and $f_2^+$) would correspond to NV B in the i-layer close to the $n^+$-edge. The innermost signals ($f_3^-$ and $f_3^+$) would come from NV B at a small electric field position in the i-layer. When we compare the splitting width in the i-layer at x=0 μm, that of the NV A target is approximately 20 MHz, while that of NV B is almost doubled, at 37 MHz.

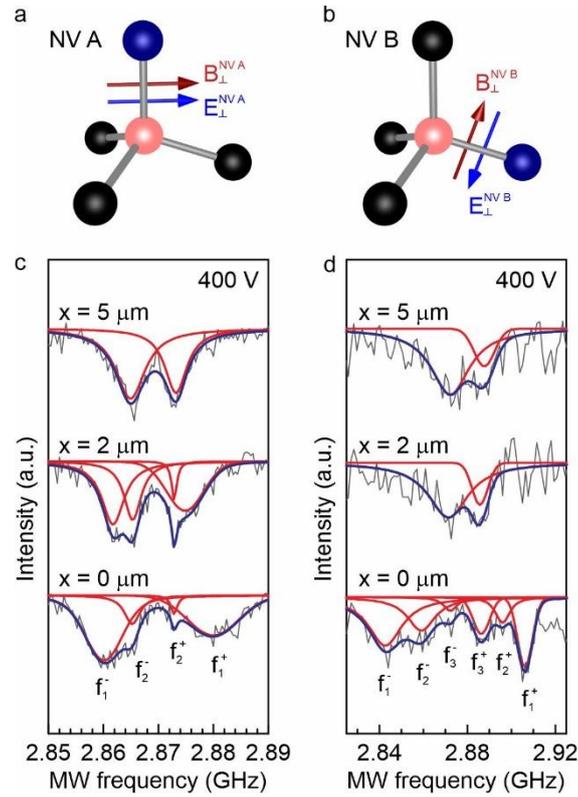

Figure 2. ODMR spectra under transverse magnetic and electric fields. Schematics of (a) NV A and (b) NV B. ODMR spectra of (c) NV A and (d) NV B at different x positions. A reverse voltage of 400 V was applied to the device to generate the electric field. The gray curve represents the experimental data. The red and blue curves are the fits and envelopes of the fitting curves, respectively.

Figure 3 shows the relationship between the increase in the ODMR splitting and the effective electric field ($\Pi_\perp$). The solid lines represent the case for equation (1). Upon application of the transverse electric field, we see nonlinear behavior at approximately zero, but it becomes a linear dependence over 0.1 MV/cm. The plots depict the experimentally obtained ODMR splitting widths including the estimates at the n$^+$-i interface; the effective electric field was estimated to range from 0.2 to 1.9 MV/cm.

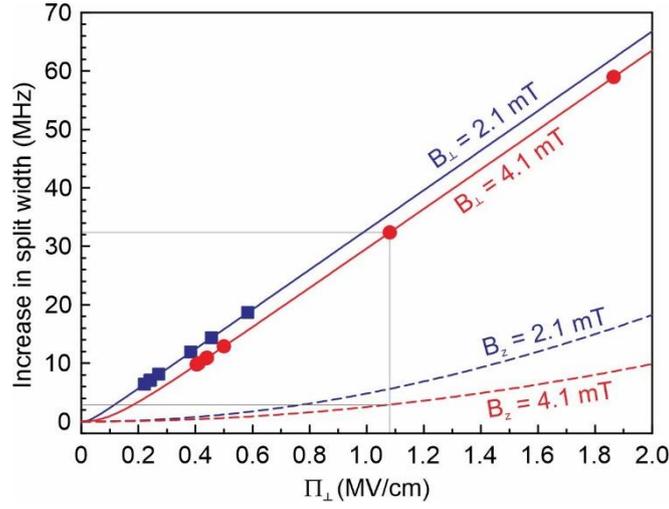

Figure 3. Increase in the split width of the ODMR dips as a function of the effective electric field. The increase in the split width is defined as the increase from the value at zero effective electric field. The lines indicate calculated values upon application of transverse (solid lines) or axial (dashed lines) magnetic fields to a target NV. The gray lines represent a comparison of the increase in split width depending on the direction of the magnetic field.

The effective electric field ($\Pi_\perp$) is composed of the electric field ($E_\perp$) and strain field ($\sigma_\perp$). Due to the unknown direction of the strain field, we consider the strain field to be an error for the estimation of the electric field. The magnitude of the strain field is obtained from the splitting of the target NV center at the zero voltage condition (Fig. 1c). An NV A splitting width of 4.6 MHz corresponds to ~0.13 MV/cm. Note that since we do not know the azimuthal angle of strain ($\phi_\sigma$) in equation (1), we took the maximum strain value at $\phi_\sigma = 0°$ as an error. The splitting of NV B is 7.9 MHz, giving rise to a strain field of ~0.2 MV/cm.

Finally, solving equation (2) for NV A and NV B, we obtained $E_x$ and $E_z$ generated in the device. These values are plotted as a function of the x position in Fig. 4b,c. As seen in Fig. 4b, the data at x>0 and x<0 correspond to the electric field in the i-layer and n$^+$-i interface, respectively. The maximum electric fields in the i-layer at x=0 μm are $E_x = 0.58 \pm 0.13$ MV/cm and $E_z = 1.35 \pm 0.26$ MV/cm. We compared the obtained values with those calculated by the device simulation, shown as lines in

Fig. 4b,c. For the device simulation, two structural models were considered. Model 1 represents a normal vertical p-i-n diode, while in model 2, an N donor distribution is added to the device, corresponding to the distribution of N ion implantation (Appendix E). The substitutional N atoms without a neighboring vacancy function as donors in diamond, with a donor level of ~1.6 eV from the minimum of the conduction band. The N atoms introduced by ion implantation are thought to take the donor state to some extent. In model 2, all implanted N atoms are assumed to be in the donor state, which forms one boundary for the simulation. Here, the effects of surface defects [14] and other nitrogen-related defects are not considered. The simulated curves represent the values at the projected depth of the N atoms in the i-layer (z=0.35 μm for x>0) and at the $n^+$-i interface (z=0 μm for x<0). Both the experimental and simulation results increase as approaching the $n^+$-edge at x=0 μm.

The experimental estimation of $E_x$ near the $n^+$-edge is within the simulation results and is close to the model that includes the effect of the N donors. The lower limits of $E_z$ corresponding to the case for the same direction of the electric and strain fields are in good agreement with the simulation. The electric fields that are expected to correspond to the $n^+$-i interface are also plotted at x=0 μm. These values are close to the simulation at z=0 μm when averaged within a few hundred nanometers in the x direction. It is worth noting that the electric fields take much higher values exactly at the $n^+$-edge in the simulation, but the number of NV centers at such a region is substantially small. Thus, the value that we obtained experimentally here should be the average in the measurement regions. Super-resolution techniques [45–47] will enable us to capture the high electric field at the concentrated region. Figure 4a depicts the electric field vectors obtained from the experimental results, clearly showing that the x component is largely generated in the i-layer at x>0 μm, while the z component is dominant at the $n^+$-i interface. Here, we estimated the components of the electric fields in the device, while we note that the experimental error due to the strain is relatively large, and thus, determination of the direction of the strain field [5] is necessary to more precisely discuss the effect of N ion implantation.

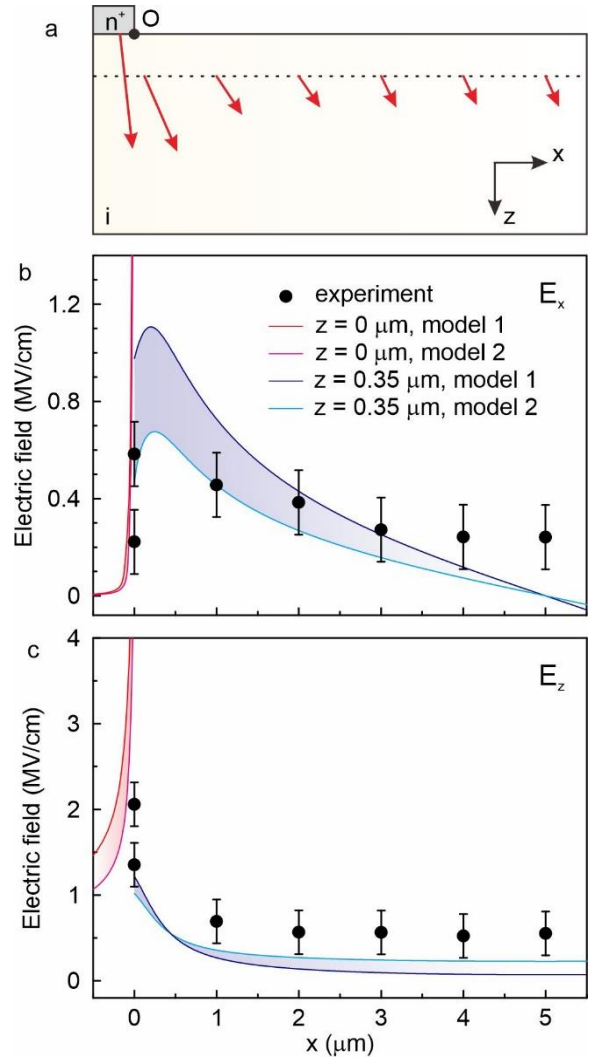

Figure 4. Comparison with simulation. (a) Schematic of the device with electric field vectors. The length of the red arrows denotes the magnitude of the electric field. The dashed line indicates z=0.35 μm. (b) $E_x$ and (c) $E_z$. The dots and lines denote the experimental and simulation results, respectively. In model 2, the implanted N atoms are treated as donors in diamond.

## IV. Conclusions

We demonstrated vector electrometry in a vertical diamond p-i-n diode using an ensemble of NV centers directly embedded in the device. Application of the transverse magnetic field to a target NV axis enabled us to identify the signals from the target in an ODMR spectrum and to enhance the ODMR response against the electric field compared with the case for application of an axial magnetic field. By performing the process for multiple NV axes, we successfully obtained the components of the electric field generated in the device and confirmed that the experimental values are close to the simulation results of a structural model that includes the implanted N donors. The vector electrometry demonstrated here will become an important technique to reveal the electric fields generated in various systems.

**Acknowledgments**

This work was supported by MEXT/JSPS KAKENHI Grant Number 18H01472, the MEXT Quantum Leap Flagship Program (MEXT Q-LEAP) Grant Number JPMXS0118067395, and the Toray Science Foundation. AY is supported by the U.S. Department of Energy, Office of Science, Office of Basic Energy Sciences Energy Frontier Research Centers program under Award Number DE-SC-0019300, the Army Research Office under Grant Number W911NF-17-1-0023 the NSF STC Center for Integrated Quantum Materials, NSF Grant No. DMR-1231319.

**Appendix A. ODMR split width under magnetic and electric fields**

Under arbitrary directed magnetic and electric fields, the ODMR splitting in an NV coordinate is described as [11]

$$\frac{1}{2}W = \left[(\gamma^2 B_z^2 + k_\perp^2 \Pi_\perp^2) - \frac{\gamma^2 B_\perp^2}{2D_{gs}}(\gamma^2 B_z^2 + k_\perp^2 \Pi_\perp^2)^{\frac{1}{2}} \sin\alpha \cos(2\phi_B + \phi_\Pi) + \frac{\gamma^4 B_\perp^4}{4D_{gs}^2}\right]^{\frac{1}{2}} \quad (3)$$

where $B_z$ is the axial magnetic field along the NV axis and $\tan\alpha = k_\perp \Pi_\perp / \gamma B_z$. $\tan\phi_B = B_y/B_x$. Note that an axial electric field leads to the center shift of the two resonance frequencies [11], and thus, no effect is observed on the splitting.

When taking $B_z \sim 0$, equation (3) becomes

$$\frac{1}{2}W = \left[k_\perp^2 \Pi_\perp^2 - \frac{\gamma^2 B_\perp^2}{2D_{gs}} k_\perp \Pi_\perp \cos(2\phi_B + \phi_\Pi) + \frac{\gamma^4 B_\perp^4}{4D_{gs}^2}\right]^{\frac{1}{2}} \quad (4)$$

Setting the direction of the transverse magnetic field along the x-axis in a target NV alignment leads to $\cos(2\phi_B + \phi_\Pi) = \cos\phi_\Pi$. Consequently, we obtain equation (1) in the main text.

On the other hand, when an axial magnetic field is applied, the first and second terms on the right-hand side, $(\gamma^2 B_z^2 + k_\perp^2 \Pi_\perp^2)$, in equation (3) become dominant, and equation (3) is reduced to

$$\frac{1}{2}W \approx [\gamma^2 B_z^2 + k_\perp^2 \Pi_\perp^2]^{\frac{1}{2}} \qquad (5)$$

This relation is illustrated as the dashed lines in Fig. 3. With a constant axial magnetic field, the increase in the split width is nonlinear against the effective electric field ($\Pi_\perp$) and becomes much lower than that for the case with a transverse magnetic field (solid lines).

**Appendix B. Measurement system**

A schematic illustration of the measurement system in the vacuum chamber is shown in Fig. 5. The sample holder is designed to be equipped with three-axis electromagnets. A large coil placed at the center and four coils with iron cores control the magnetic field along the z-axis and on the xy plane in the lab frame, respectively. We determined the direction of the transverse magnetic field by observing the smallest splitting from the target NV signals in the ODMR spectrum. The diamond devices have electrical contacts to apply a bias. The MW radiation for electron spin resonance of the NV centers is applied via a thin Cu wire. The microwave wire, electromagnets, and electrical contacts are connected to controllers outside the chamber through feedthrough connectors.

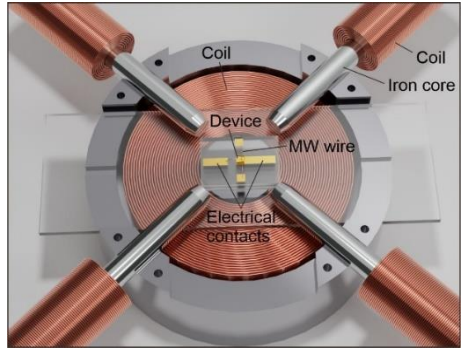

Figure 5. Schematic of measurement system in the vacuum chamber.

**Appendix C. Unit vectors**

The unit vector for each NV alignment in the lab frame defined in Fig. 1d is given as

$$\mathbf{u}_{NV\,A} = \begin{pmatrix} 0 \\ 0 \\ -1 \end{pmatrix}$$

$$\mathbf{u}_{NV\,B} = \begin{pmatrix} \sin\theta \\ 0 \\ -\cos\theta \end{pmatrix}$$

$$\mathbf{u}_{NV\,C} = \begin{pmatrix} \sin\theta \cos 120° \\ \sin\theta \sin 120° \\ -\cos\theta \end{pmatrix}$$

$$\mathbf{u_{NV\,D}} = \begin{pmatrix} \sin\theta \cos 240° \\ \sin\theta \sin 240° \\ -\cos\theta \end{pmatrix}$$

where $\theta = 109.47°$ corresponds to the angle between the N-V axes. By substituting into equation (2) and solving a system of equations, the components of the electric field ($E_x$, $E_y$, and $E_z$) can be calculated. $E_y$ is negligible in the device used in this study. Thus, we estimated $E_x$ and $E_z$ using the two equations regarding NV A and NV B.

**Appendix D. Electrical characterization**

Figure 6a shows an I-V curve of the vertical diamond p-i-n diode incorporating an ensemble of NV centers. The current rapidly increases at a forward voltage of approximately -4 V, corresponding to the built-in potential of diamond. The device shows a high rectification ratio of ~$10^6$ at $\pm 10$ V, indicating good diode operation. The breakdown does not occur even at a high voltage over 400 V (Fig. 6b). The reverse current of devices at 100 V with different amounts of NV centers is summarized in Fig. 6c. Except for one clean device without the NV center (#3), the reverse current is ~$10^{-7}$ A. Thus, NV fabrication does not affect the reverse current. Device #3 might possess defects in the device that cause the leakage path [48].

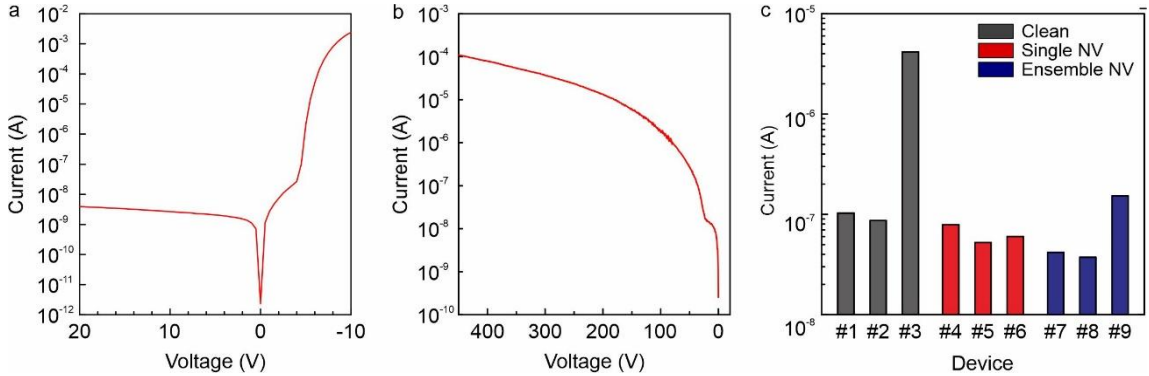

Figure 6. Electrical characteristics of diamond p-i-n diodes. (a) I-V curve of a device with ensemble NV centers, measured using a vacuum probe system. (b) High-voltage property, measured in the vacuum chamber for electric field sensing. (c) Current at a reverse bias of 100 V of devices with different NV amounts, measured in a vacuum probe system. Clean, Single NV, and Ensemble NV denote the device without N ion implantation, with a N dose of $1\times 10^9$ cm$^{-2}$ and with a N dose of $1\times 10^{12}$ cm$^{-2}$, respectively.

**Appendix E. Simulation model**

Figure 7 depicts the two structural models used for the device simulation. Model 1 is a normal p-i-n diode. The n$^+$ regions on top have a phosphorus concentration of $1\times 10^{19}$ cm$^{-3}$, and the p substrate is boron-doped, with a concentration of $1\times10^{17}$ cm$^{-3}$. The boron concentration of the i-layer is $2\times10^{14}$ cm$^{-3}$, which is the detection limit of secondary ion mass spectrometry. In model 2, the nitrogen donors are included in the i-layer and at the n$^+$-i interface, with a projected depth of 350 nm. These two models define the boundaries of the electric field simulation, and the values with intermediate nitrogen concentrations are between the two models. We found that in model 2, $E_x$ takes negative values at low voltages depending on the nitrogen concentration, and then, it becomes positive as the voltage increases. This finding explains why the $E_x$ values in model 2 are lower than those in model 1 in Fig. 4b. As expected, however, application of a voltage of 400 V leads to a positive $E_x$ value.

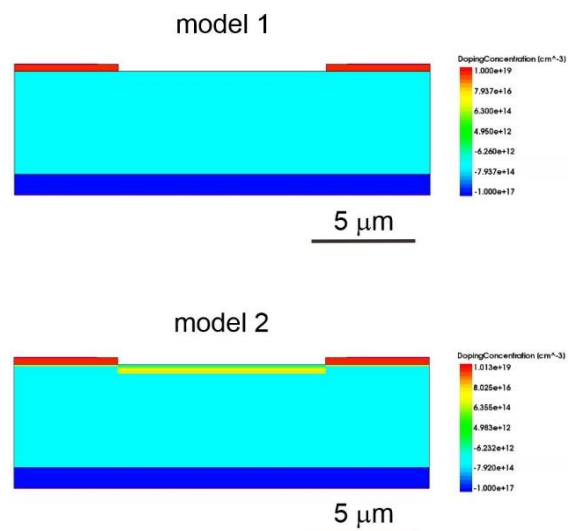

Figure 7. Structural models in the simulation, showing the doping concentrations. The positive and negative values correspond to the donor and acceptor concentrations, respectively.